\documentclass{aa}
\usepackage[varg]{txfonts}

\usepackage{epsfig,graphicx,natbib,url,twoopt}
\usepackage{enumitem}
\usepackage[
  breaklinks = true,
  colorlinks = true,
  urlcolor   = blue,
  citecolor  = blue,
  linkcolor  = blue,
]{hyperref}
\usepackage{float}
\setcitestyle{citesep={,}} 

\def\Halpha{\mbox{H\hspace{0.1ex}$\alpha$}}
\def\Hbeta{\mbox{H\hspace{0.1ex}$\beta$}}
\def\Hepsilon{\mbox{H\hspace{0.1ex}$\varepsilon$}}

\def\FeII{\ion{Fe}{ii}}

\def\NiII{\ion{Ni}{ii}}

\def\MgK{\ion{Mg}{ii}~k}

\def\Mgtriplet{\ion{Mg}{ii}~triplet}
\def\Siiv{\ion{Si}{iv}}

\defcitealias{2024A&A...689A.156B}{Paper I}

\begin{document} 
\title{Magnetic Topology of quiet-Sun Ellerman bombs and associated Ultraviolet brightenings}
\titlerunning{Magnetic Topology of QSEBs and associated UV brightenings}

\author{Aditi Bhatnagar \inst{1,2}
\and Avijeet Prasad \inst{1,2}
\and Luc Rouppe van der Voort \inst{1,2}
\and Daniel Nóbrega-Siverio \inst{3,4,1,2}
\and Jayant Joshi \inst{5}}      

\institute{
  Institute of Theoretical Astrophysics,
  University of Oslo, %
  P.O. Box 1029 Blindern, N-0315 Oslo, Norway
\and
  Rosseland Centre for Solar Physics,
  University of Oslo, %
  P.O. Box 1029 Blindern, N-0315 Oslo, Norway
\and
  Instituto de Astrofísica de Canarias, 
  38205 La Laguna, Tenerife, Spain
\and
  Universidad de La Laguna, Dept. Astrofísica, 
  38206 La Laguna, Tenerife, Spain
\and
  Indian Institute of Astrophysics, 
  II Block, Koramangala, Bengaluru 560 034, India
}


 
\date{submitted to A\&A Oct 30, 2024 / accepted Dec 7, 2024} 

\abstract
{ 
Quiet-Sun Ellerman bombs (QSEBs) are small-scale magnetic reconnection events in the lower atmosphere of the quiet Sun. Recent work has shown that a small percentage of them can occur co-spatially and co-temporally to ultraviolet (UV) brightenings in the transition region.}
{ 
We aim to understand how the magnetic topologies associated with closely occurring QSEBs and UV brightenings can facilitate energy transport and connect these events.}
{ 
We used high-resolution \Hbeta\ observations from the Swedish 1-m Solar Telescope (SST) and detected QSEBs using \textit{k}-means clustering. We obtained the magnetic field topology from potential field extrapolations using spectro-polarimetric data in the photospheric \ion{Fe}{i} 6173~\AA\ line.
To detect UV brightenings, we used coordinated and co-aligned data from the Interface Region Imaging Spectrograph (IRIS) and imposed a threshold of 5$\sigma$ above the median background on the (IRIS) 1400~\AA\ slit-jaw image channel.}
{ 
We identify four distinct magnetic configurations that associate QSEBs with UV brightenings, including a simple dipole configuration and more complex fan-spine topologies with a three-dimensional (3D) magnetic null point. In the fan-spine topology, the UV brightenings occur near the 3D null point, while QSEBs can be found close to the footpoints of the outer spine, the inner spine, and the fan surface. We find that the height of the 3D null varies between 0.2~Mm to 2.6~Mm, depending on the magnetic field strength in the region. We note that some QSEBs and UV brightenings, though occurring close to each other, are not topologically connected with the same reconnection process. We find that the energy released during QSEBs falls in the range of $10^{23}$ to $10^{24}$~ergs.}
{ 
This study shows that magnetic connectivity and topological features, like 3D null points, are crucial in linking QSEBs in the lower atmosphere with UV brightenings in the transition region.}


\maketitle

\section{Introduction}
\label{sec:introduction}
Ellerman Bombs (EBs) are small-scale, short-lived enhancements observed in the wings of the \Halpha\ spectral line at 6563 \AA\, first reported by \citet{1917ApJ....46..298E}.
These events typically occur in solar active regions. They are driven by magnetic reconnection in the lower solar atmosphere, particularly in the photosphere, due to interactions between opposite magnetic polarities during magnetic flux emergence. 
EBs are distinguished by their characteristic moustache-like spectral profile \citep{1964ARA&A...2..363S}, where the \Halpha\ wings show enhanced emission while the core remains unperturbed in absorption. 
When observed near the solar limb, EBs appear as tiny bright flames, with lifetimes ranging from a few seconds to several minutes \citep[e.g.,][]{1982SoPh...79...77K, 1998SoPh..182..381N, 2011ApJ...736...71W, 2013JPhCS.440a2007R, 2015ApJ...798...19N}. 
Recently, events similar to EBs were also discovered in quieter parts of the Sun, away from active regions, by \citet{2016A&A...592A.100R}, using \Halpha\ observations from the Swedish 1-m Solar Telescope \citep[SST,][]{2003SPIE.4853..341S}. 
These are called Quiet Sun Ellerman Bombs (QSEBs). They share many characteristics with EBs but occur in less magnetically active areas.
\citet{2020A&A...641L...5J} and \citet{2022A&A...664A..72J} found that QSEBs are quite ubiquitous, estimating that around 500\,000 QSEBs occur on the Sun at any given moment using high-resolution \Hbeta\ observations from SST. 
This estimate was revised to 750\,000 by \citet{2024A&A...683A.190R}, who utilised \Hepsilon\ observations for improved detection due to the higher spatial resolution and larger contrast of the shorter wavelength \Hepsilon\ line.
\begin{figure*}[!ht]
    \centering
    \includegraphics[width=\linewidth]{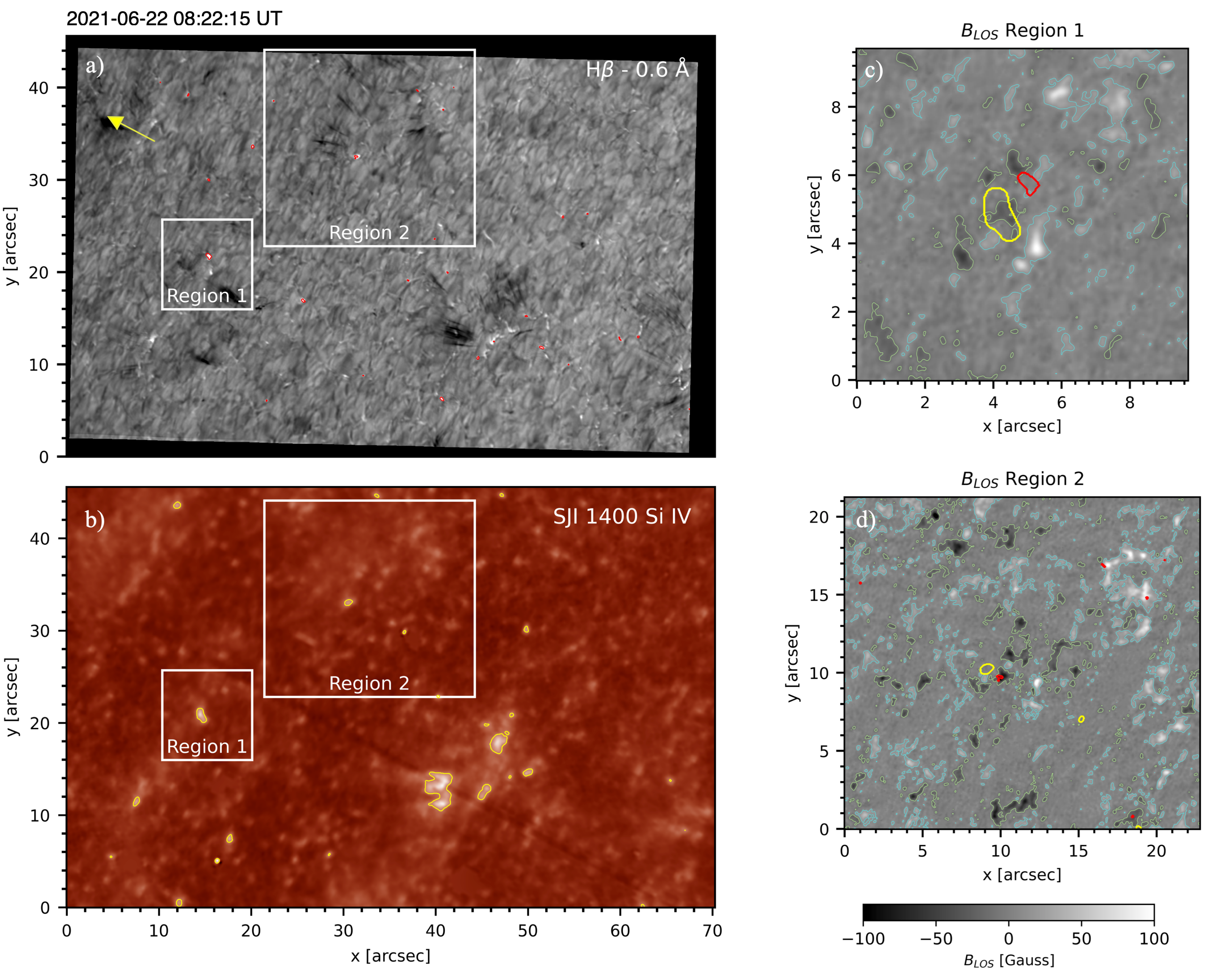}
    \caption{Overview of the observed region in \Hbeta\ blue wing, SJI~1400, and magnetic field ($B_\mathrm{LOS}$). The two white squares mark the regions with continuous QSEB and UV brightening activity. The first four events analysed in this paper occur in Region~1, and the fifth event occurs in Region~2. Red contours at the top outline the QSEB detections. Yellow contours outline $>$5$\sigma$ UV brightenings. The arrow in the top panel shows the direction towards the nearest limb. $B_\mathrm{LOS}$ maps with contours at 2$\sigma$ above the noise level for these two regions are shown at right, with the QSEB and UV brightening contours. The values for $B_\mathrm{LOS}$ in panels c) and d) have been saturated to $\pm$ 100 G for better representation.}
    \label{fig:1}
\end{figure*}

Both QSEBs and EBs occur due to magnetic reconnection, and several studies have explored various topological scenarios which can explain the occurrence of EBs. 
One of the scenarios is the magnetic reconnection between the newly emerging flux and pre-existing magnetic fields \citep[e.g.,][]{2008ApJ...684..736W, 2010PASJ...62..879H, 2017ApJ...839...22H, 2024A&A...686A.218N}. 
In another scenario, EBs can arise in unipolar regions due to misalignment of field lines, causing the formation of a quasi-separatrix layer \citep[QSL,][]{1996A&A...308..643D} between regions with the same magnetic polarity. 
There, shearing between the different topological configurations can trigger reconnection \citep{2002ApJ...575..506G, 2008ApJ...684..736W, 2010PASJ...62..879H}. 
Another scenario proposed by \citet{2004ApJ...614.1099P, 2006AdSpR..38..902P, 2012ASPC..455..177P, 2012EAS....55..115P} suggests that EBs can occur at bald patches, where U-shaped photospheric magnetic loops are prone to reconnection. They can also happen at the separatrices associated with bald patches in emerging serpentine-shaped magnetic fields.
These different reconnection scenarios highlight the complexity and diversity of magnetic topologies that can lead to the formation of EBs. It is, therefore, worth investigating if this complexity in the magnetic topology also extends to QSEBs.

\begin{figure*}
    \centering
    \includegraphics[width=\linewidth]{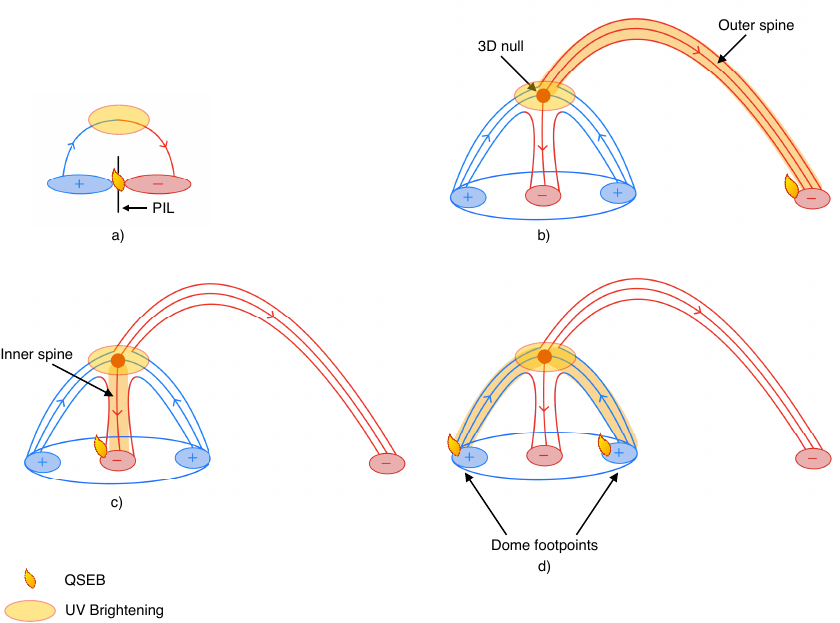}
    \caption{Cartoons schematising the magnetic topologies derived from potential field extrapolations for co-occurring QSEBs and UV brightenings. Panel (a) shows a dipole configuration with a black vertical line marking the polarity inversion line (PIL). Panels (b), (c), and (d) depict UV brightenings occurring at the null point of a fan-spine topology. These three panels highlight various possible locations for QSEBs: near the footpoint of the outer spine (b), near the footpoint of the inner spine (c), and near the dome footpoints (d). The orange-shaded parts highlight the possible regions influenced by the energy flow from the reconnection site.}
    \label{fig:2}
\end{figure*}

Higher in the solar atmosphere, regions with opposite magnetic polarities can host ultraviolet (UV) bursts. 
They are observed with the Interface Region Imaging Spectrograph \citep[IRIS,][]{2014SoPh..289.2733D} 
as compact, intense and rapidly varying brightenings in the \Siiv\ spectral lines \citep{2014Sci...346C.315P}.
These events are typically smaller than 2\arcsec\ and have durations ranging from mere seconds to over an hour \citep{2018SSRv..214..120Y}.
The \Siiv\ emission lines observed during UV bursts are often broadened with their spectral wings, including absorption blends from neutral or singly ionised species such as \FeII\ and \NiII. 
This suggests that the plasma, heated to transition region temperatures of around 100,000 K, is located beneath a cooler chromospheric canopy of fibrils \citep{2014Sci...346C.315P}, an interpretation that has been supported by several studies since then \citep[e.g.,][]{2015ApJ...812...11V, 2015ApJ...809...82G, 2017MNRAS.464.1753H, 2022A&A...657A.132K}.
Like EBs, UV bursts are also triggered by magnetic reconnection in complex topological configurations in the solar atmosphere. 
\cite{2017ApJ...836...63T} and \citet{2017ApJ...836...52Z} demonstrated that UV bursts frequently appear at bald patches in emerging flux regions. 
On the other hand, \citet{2018ApJ...854..174T} found in their observations that only a small portion of UV bursts could be linked to bald patches.
Most of these events take place in regions with high squashing factors \citep{1996A&A...308..643D, 2002JGRA..107.1164T, 2005LRSP....2....7L}, around 1 Mm above the surface, where strong variations in magnetic connectivity and electric currents are present. This is similar to the mechanism observed for EBs \citep{2004ApJ...614.1099P}. 
Studies by \citet{2017A&A...605A..49C} and \citet{2018A&A...617A.128S} have highlighted the fan-spine topology as an apt magnetic configuration for UV burst generation. 
Fan-spine topologies can also form in case of flux emergence, with a null point present somewhere in the atmosphere. 
\citet{2017ApJ...851L...6R} and \citet{2017ApJ...850..153N} show that UV bursts occur as a result of magnetic reconnection between newly emerging magnetic domes and the pre-existing ambient magnetic field. 
This structure consists of a dome-shaped fan surface with a three-dimensional (3D) magnetic null point where the magnetic field strength goes to zero \citep{2002A&ARv..10..313P, 2005LRSP....2....7L}. 
The inner and outer spine pass through this null point, with their footpoints rooted in regions with the same magnetic polarity, while the fan surface (or dome) footpoints are anchored in a ring-shaped area of opposite polarity around the inner spine. 
The null point acts as a potential site for magnetic reconnection, facilitating energy release and triggering a UV burst at this location.

\begin{figure*}
    \centering
    \includegraphics[width=\linewidth]{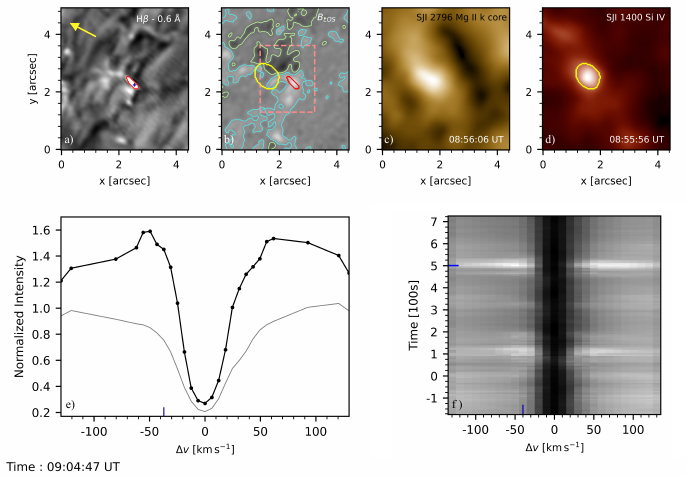}
    \caption{Details of the QSEB with dipole magnetic topology in Region~1, at the instance of maximum wing enhancement in the wings of the \Hbeta\ line. The top row shows the QSEB in \Hbeta\ $-0.6$~\AA, the $B_\mathrm{LOS}$ map with contours at 2$\sigma$ above the noise level, and SJI 2796 and 1400. SJI 2796 and 1400 are at a different time, as the UV brightening begins before the QSEB. The yellow contour in panel d) shows a region with $>$5$\sigma$ intensity in SJI~1400. The pink dashed box in panel b) encloses the area considered to calculate the magnetic flux and magnetic energy shown in Fig.~\ref{fig:5}. Panel e) shows the normalised intensity in \Hbeta\ at the blue cross in the above \Hbeta\ $-0.6$~\AA\ image. Panel f) shows the $\lambda t$ diagram for the blue marker over the QSEB. The position of this marker has been adjusted a few times to follow the strongest EB spectra for each timestep. The vertical blue markers in panels e) and f) denote the position of the wing of \Hbeta\ line for which the image is shown in panel a). The horizontal blue marker in panel f) denotes the time of the snapshot shown in this figure. An animation of this figure, which shows the evolution of both the QSEB events discussed in \ref{sec: 4.1}, is available in the online material (see \url{http://tsih3.uio.no/lapalma/subl/qseb_uvb_topology/movies/bhatnagar_movie_fig03.mp4}).}
    \label{fig:3}
\end{figure*}
%
\citet{2015ApJ...812...11V} demonstrated simultaneous occurrences of EBs and UV bursts, observed through IRIS spectra in the \Mgtriplet\ and \Siiv\ lines. Their analysis showed that EBs could heat plasma to transition region temperatures much higher than previously thought.
Similarly, \citet{2016ApJ...824...96T} identified 10 UV bursts associated with EBs, showing notable \Halpha\ wing brightening but lacking a line core signature, indicating that these events might share similar origins.
\citet{2019A&A...626A..33H} used a 3D radiative magnetohydrodynamic (MHD) Bifrost simulation \citep{2011A&A...531A.154G} to model how EBs and UV bursts can form along an extended current sheet at different heights in the solar atmosphere. 
They found that EBs and UV bursts are part of the same magnetic reconnection system, occurring at different heights along the current sheet. 
Specifically, EBs are confined to the lower photosphere up to around 1200~km, while UV bursts form at higher altitudes, ranging from 700~km to 3~Mm above the photosphere. 
The study suggests that the orientation of the current sheet and the viewing angle can cause spatial offsets between EB and UV burst occurrences.
\citet{2023A&A...672A..47S} used MHD simulations with the MURaM \citep{2005A&A...429..335V, 2017ApJ...834...10R, 2022A&A...664A..91P} code to explore small-scale loop-like structures in the solar atmosphere. 
The synthetic \Halpha\ and \Siiv\ signatures in their simulations have been linked to EBs and UV bursts, with some events showing coincident emissions in both observables.
This is further supported by the observations of 
\citet{2015ApJ...812...11V}, 
\citet{2019ApJ...875L..30C}, and
\citet{2020A&A...633A..58O}, which demonstrates that UV bursts frequently co-occur close to EBs with a spatial offset towards the limb.
\citet{2017ApJ...845...16N} provided the first evidence of transition region response to a QSEB, detecting \Siiv\ emissions using coordinated \Halpha\ and IRIS observations.

In our previous study, \citet[][hereafter \citetalias{2024A&A...689A.156B}]{2024A&A...689A.156B}, we investigated the spatial and temporal relationship between QSEBs and UV emissions in IRIS observations. 
We found that the \Siiv\ emission did not result in as intense and extremely broadened spectral profiles as in active region UV bursts and referred to these events as UV brightenings. 
We also showed that 15\% of long-lived ($>1$~min) QSEBs could be associated with UV brightenings in the \Siiv\ lines. 
This analysis revealed that QSEBs often precede UV brightenings, which occur within 1000~km of the QSEB, typically in the direction of the solar limb. 
Some QSEBs, which were sampled by the IRIS slit, also exhibited emissions in the \Siiv\ and \Mgtriplet\ spectral lines, implying that QSEBs can generate localised heating up to transition region temperatures.
The observations of \citetalias{2024A&A...689A.156B} 
include high-quality measurements of the photospheric magnetic field. In this paper, we build upon this foundation and analyse the magnetic topology of regions with co-spatial and co-temporal QSEBs and UV brightenings, as well as explore different reconnection scenarios that can relate these events.

\begin{figure*}
    \centering
    \includegraphics[width=\linewidth]{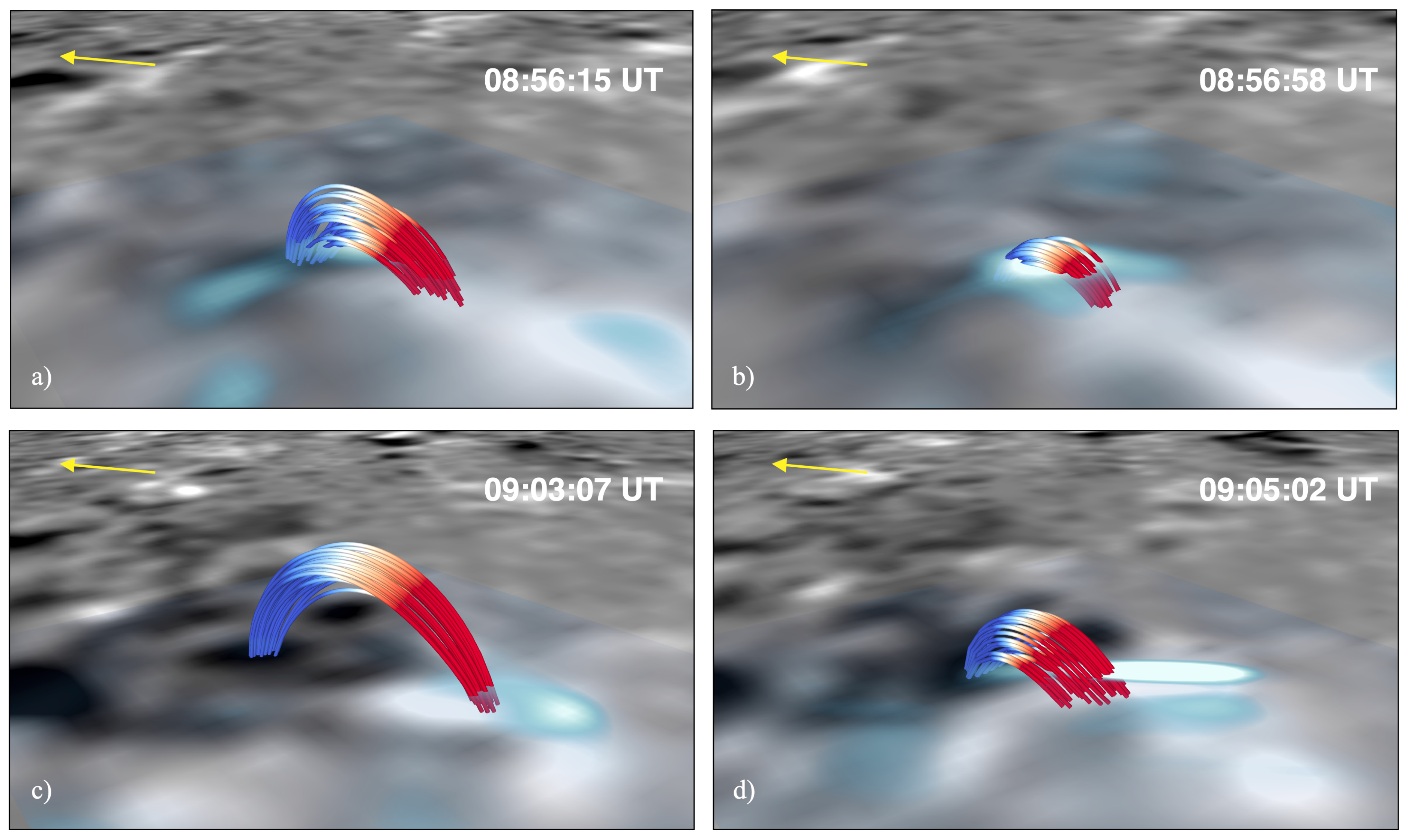}
    \caption{Magnetic topology of the two QSEBs with dipole configuration in Region~1. Panels a) and c) show the loops connecting the positive and negative polarities before the two QSEBs occur. The height of the loops in panels a) and c) are 200~km and 323~km, respectively. The heights of the same loops reduce to 90~km and 160~km in panels b) and d) when the QSEBs occur. In this (and subsequent) VAPOR visualisation(s), the grey-scale image at the bottom shows the $B_\mathrm{LOS}$ map, and the higher image shows the \Hbeta\ wing in light blue and white shades. The QSEBs are visible as elongated white patches in the right column. The yellow arrow shows the direction towards the closest solar limb.}
    \label{fig:4}
\end{figure*}

\section{Observations}
\label{sec:observations}

The observations are a 51~min time series acquired with the SST in quiet Sun near the North limb ($\mu=0.48$) on 22 June 2021 from 08:17:52 UT to 09:08:32 UT. 
The CHROMIS instrument \citep{Scharmer2017} 
acquired $\Hbeta$ spectral data at 27 line positions at a temporal cadence of 7~s. 
The CRISP instrument \citep{2008ApJ...689L..69S} 
provided spectro-polarimetric data in the \ion{Fe}{i} 6173~\AA\ line at a cadence of 19~s. 
Milne-Eddington inversions were performed to estimate the full magnetic field vector using an inversion code by \cite{2019A&A...631A.153D}. 
The noise level in the maps of the line of sight magnetic field ($B_{\mathrm{LOS}}$) was estimated to be 6.4~G, as done in \citetalias{2024A&A...689A.156B}.
High data quality was achieved with the aid of the SST adaptive optics system \citep{2024A&A...685A..32S} 
and image restoration using multi-object multi-frame blind deconvolution \citep[MOMFBD,][]{2005SoPh..228..191V} as part of the standard SSTRED data reduction pipeline
\citep[]{2015A&A...573A..40D, 2021A&A...653A..68L}.

IRIS observed the same region as part of a coordinated observation campaign with SST.
Here we use the slit jaw images (SJI) in the 1400~\AA\ and 2769~\AA\ channels that were acquired at a cadence of 18~s. 
The SJI~1400~\AA\ channel is dominated by emission in the transition region \Siiv~1394~\AA\ and 1403~\AA\ spectral lines. 
The SJI~2769~\AA\ channel is centred on the \MgK\ line core.  We aligned observations from the Atmosphere Imaging Assembly 
\citep[AIA,][]{2012SoPh..275...17L} 
instrument on board the Solar Dynamics Observatory 
\citep[SDO,][]{2012SoPh..275....3P} 
to the SST observations following a procedure developed by Rob Rutten\footnote{\url{https://robrutten.nl/Recipes_IDL.html}}.
For more details on the observations and alignment between the different spectral lines and channels, we refer to \citetalias{2024A&A...689A.156B}. 

\begin{figure}
    \centering
    \includegraphics[width=\linewidth]{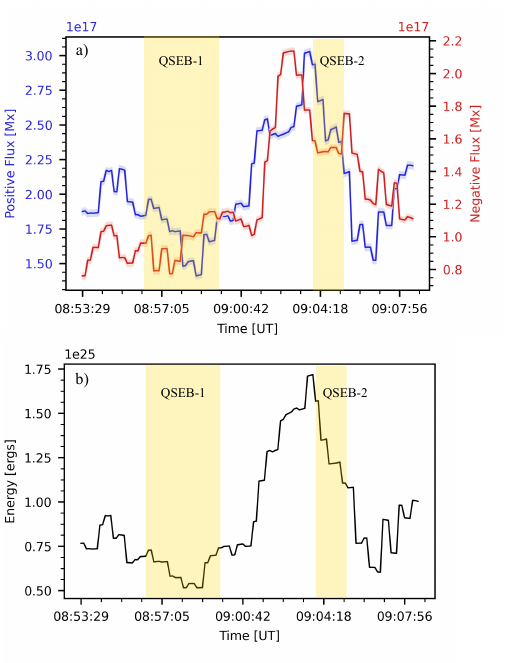}
    \caption{Evolution of positive and negative magnetic flux and magnetic energy for the two QSEBs with dipole magnetic field configuration in Region~1. The error in positive and negative flux is shown as thin-shaded regions along the curves. The region used to calculate these quantities is outlined in Fig.~\ref{fig:3}b. The yellow-shaded regions denote the QSEB occurrences.}
    \label{fig:5}
\end{figure}

\section{Method of analysis}
\label{sec:methods}

\subsection{Identification of events}
A detailed description of the method of identifying QSEB events in the \Hbeta\ data, the associated UV brightenings in the SJI~1400 data, and steps for linking them together is given in \citetalias{2024A&A...689A.156B}. 
In short, the QSEB detection employs \textit{k}-means clustering \citep{Everitt_1972} to identify characteristic EB spectral profiles and subsequently uses connected component analysis to connect EB profiles in space and time. 
Each QSEB event received an event ID number and was tracked using the Trackpy Python library\footnote{\url{https://soft-matter.github.io/trackpy/v0.6.4/}}.
In total, 1423 QSEB events were detected in the 51~min time series. 
For the UV brightenings, we focused on the strongest events by employing a threshold of 5$\sigma$ above the median background. 
This resulted in the detection of 1978 UV brightening events. 
We found that many of the associated UV brightenings occur within 1000~km of the QSEBs.
%
\begin{figure}
    \centering
    \includegraphics[width=\linewidth]{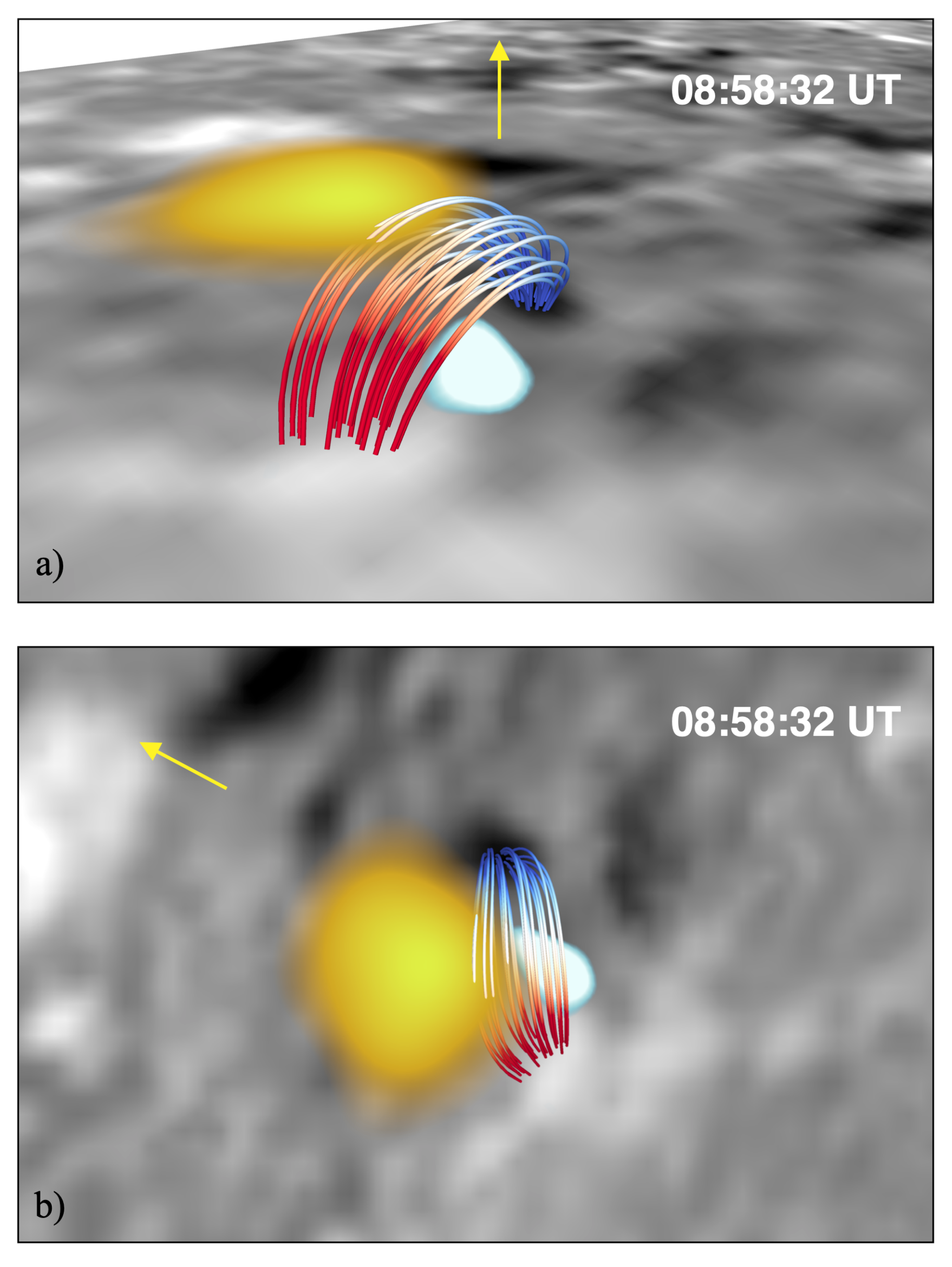}
    \caption{Magnetic topology for a dipole configuration showing the co-spatial and co-temporal QSEB and UV brightening in Region~1. Panel a) presents a 3D view with the north limb pointing up, with the QSEB close to the photosphere and the UV brightening at the top of the loops, slightly offset towards the north limb. Like in Fig.~\ref{fig:4}, the grey-scale image in the bottom is the $B_\mathrm{LOS}$ map, and the higher image is the \Hbeta\ wing image with the QSEB in light blue and white shades. In this (and subsequent) 3D visualisation(s), the UV brightening in SJI~1400 is shown as a yellow patch. Panel b) shows a top view of the loops, QSEB and UV brightening. The yellow arrow shows the direction towards the limb.}
    \label{fig:6}
\end{figure}

This paper focuses on two regions where QSEBs and UV brightenings were repeatedly observed throughout the time series. 
The regions labelled as Region~1 and Region~2 are marked in Fig.~\ref{fig:1}.
The figure shows an instance when strong QSEB events were visible in the \Hbeta\ wing and their associated $>$5$\sigma$ UV brightenings visible in SJI~1400 in these two regions. 
From Region~1, we examine four instances of QSEBs occurring at the same location at different times during the time series, and from Region~2, we study two QSEBs occurring at the same location and at different times. 
These are chosen to highlight the varied magnetic field configurations at the site where the QSEBs and UV brightenings occur. 
The UV brightenings in Region~1 are at spatial offsets ranging between 390~km and 980~km from the QSEB in the direction of the closest limb. 
For Region~2, The UV brightenings also occur with some spatial offset towards the limb at distances between 560~km and 740~km, from the QSEB.

\subsection{Magnetic field extrapolation}
We perform Fast Fourier Transformation (FFT) based potential field extrapolations \citep{1972SoPh...25..127N, 1981A&A...100..197A} on both the regions using the corresponding $B_\mathrm{LOS}$ data. The box sizes for the extrapolation were 256 $\times$ 256 $\times$ 256 pixels for Region~1 and 600 $\times$ 560 $\times$ 256 pixels for Region~2. This corresponds to a physical domain size of 7~Mm $\times$ 7~Mm $\times$ 7~Mm for Region~1 and  16.5~Mm $\times$ 15.4~Mm $\times$ 7~Mm for Region~2. The regions were chosen such that the bottom boundary was approximately in flux balance when performing the magnetic field extrapolations. This ensured that the resultant extrapolated magnetic field approximately satisfied the divergence-free condition. The mean of $B_\mathrm{LOS}$ for Regions~1 and 2 are 0.6~G and 2.0~G, respectively, which are well below the 6.4~G noise level in $B_\mathrm{LOS}$. The corresponding values for the ratio of total flux to the total unsigned flux are 0.09 and 0.25, respectively. 
The magnetic field lines for the extrapolated field were drawn using the visualisation software VAPOR \citep{2019Atmos..10..488L}.
The magnetic field lines shown in the subsequent figures were traced by selecting a small region near the base of the QSEB and randomly placing the seed points with a bias towards stronger values of absolute $B_\mathrm{LOS}$ to initiate the bidirectional field line integration. This allows us to continuously follow the strong polarities as they move during the time series. To detect the location of the magnetic null points, we biased the seed points with a large squashing factor \citep{2002JGRA..107.1164T, 2009ApJ...693.1029T}. This allowed us to follow the changes in fan-spine topology associated with the 3D null points during the event in more detail. The squashing factor has been computed with the code by \citet{2016ApJ...818..148L}. For a visual comparison of the extrapolated magnetic field lines with QSEBs in \Hbeta\ and UV brightenings in the SJI~1400~\AA\ channel in 3D, we have placed the QSEBs in \Hbeta\ slightly above the photosphere, while for the UV brightenings, the SJI~1400 layer is placed at different heights based on the height of the 3D null point.

\begin{figure*}
    \centering
    \includegraphics[width=\linewidth]{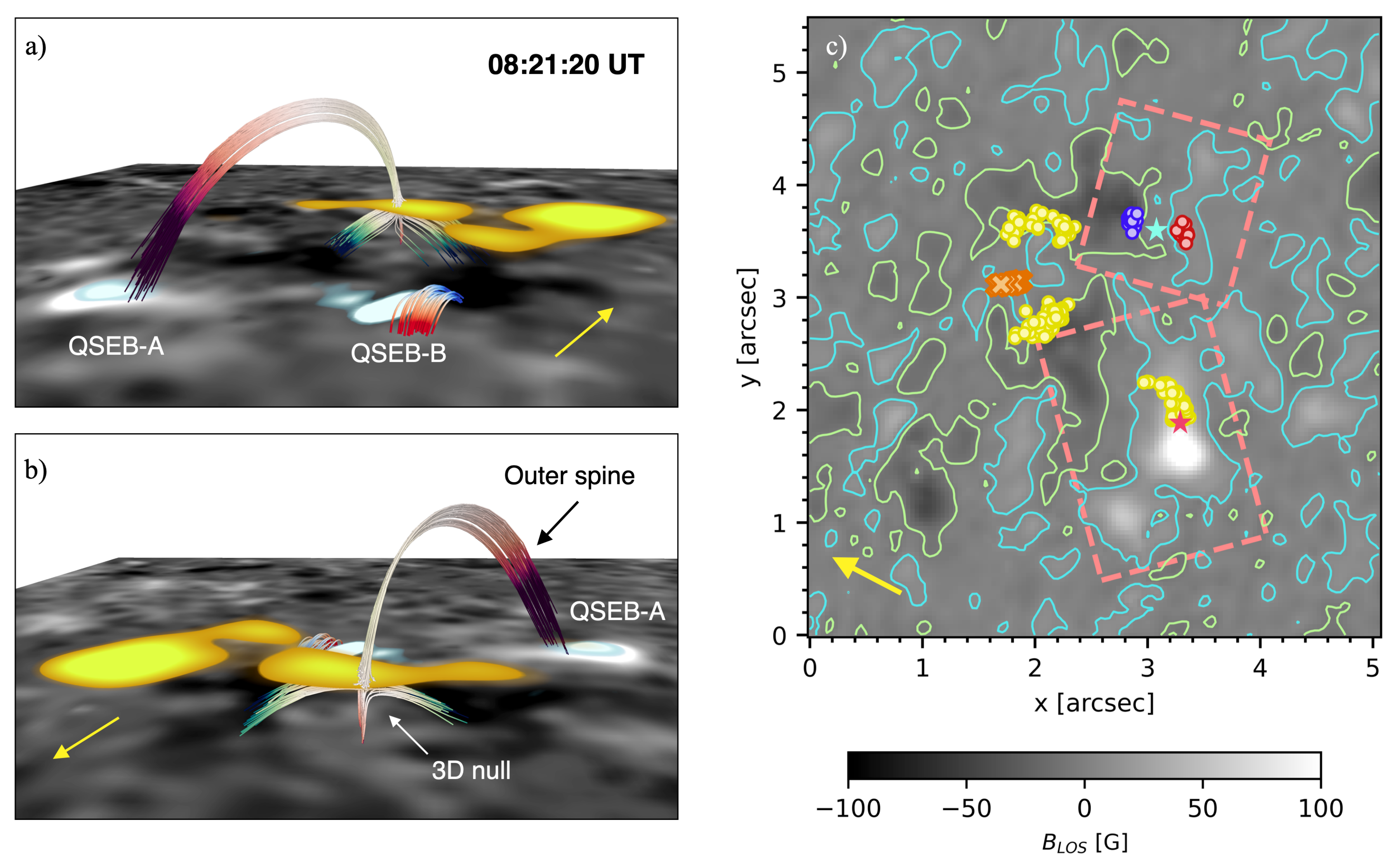}
    \caption{Magnetic fan-spine topology showing a UV brightening at a magnetic null point and a QSEB at the outer spine's footpoint, resembling the configuration in Fig.~\ref{fig:2}b. Panels a) and b) show the same instance from different viewpoints. Two QSEBs are observed near the UV brightening: QSEB-A is positioned at the footpoint of the outer spine, while QSEB-B is located at a dipole configuration, with one footpoint on the negative polarity patch where the footpoint of the fan-spine with 3D null is also located. QSEB-B is not part of the same reconnection event as the UV brightening and QSEB-A, which are associated with the fan-spine topology. Panel c) shows the $B_\mathrm{LOS}$ map with contours at 1$\sigma$ above the noise level with different markers: The orange crosses denote the projection of the null point of the UV brightening, and yellow circles denote the footpoints of the dome. The yellow circles on the positive polarity show the footpoints of the outer spine. The red star marker denotes where QSEB-A is located. The blue and red circles denote the footpoints of the dipole of QSEB-B. The cyan star points to the location of QSEB-B, which occurs at the PIL. The pink dashed boxes show the area considered to calculate the magnetic flux and magnetic energy for the two QSEBs, as shown in Fig.~\ref{fig:8}. Yellow arrows in all panels show the direction towards the limb.}
    \label{fig:7}
\end{figure*}

\begin{figure*}
    \centering
    \includegraphics[width=\linewidth]{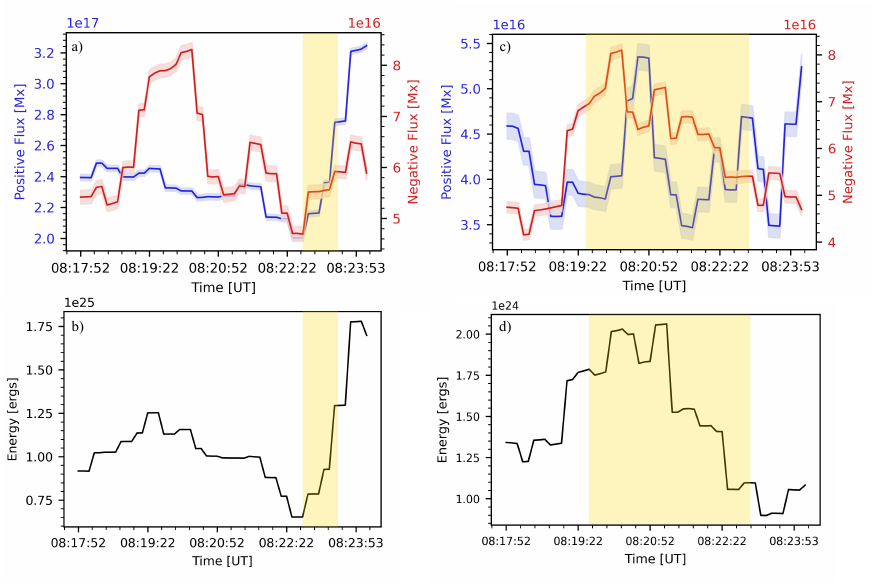}
    \caption{Evolution of positive and negative magnetic flux and magnetic energy for the two QSEBs of Fig.~\ref{fig:7}. The error bands in positive and negative fluxes are shown as thin-shaded regions along the curves. Panels a) and b) are for QSEB-A, which is located at the outer spine footpoint. Panels c) and d) are for QSEB-B, which is located in a dipole configuration. The yellow-shaded regions highlight the duration of the two QSEB events.}
    \label{fig:8}
\end{figure*}
\begin{figure*}
    \centering
    \includegraphics[width=\linewidth]{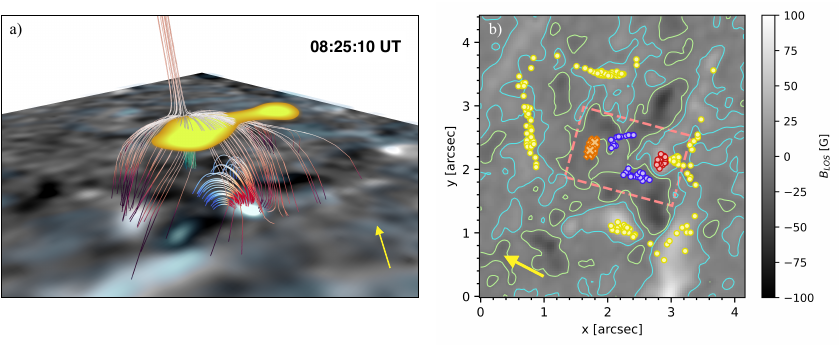}
    \caption{Magnetic fan-spine topology showing the UV brightening at the 3D null and QSEB at one of the footpoints of the fan surface. This resembles the cartoon shown in Fig.~\ref{fig:2} d). The QSEB also has a dipole configuration. The $B_\mathrm{LOS}$ map shows the footpoints of these two phenomena: orange crosses denote the projection of the null point of the UV brightening, and yellow circles denote the footpoints of the fan surface. Blue and red markers are the negative and positive footpoints of the dipole of the QSEB, respectively. The QSEB and UV brightening share a common negative polarity, and the QSEB occurs at the positive polarity where the footpoints of the fan surface are located (intersection of red and yellow dots around x, y = 3\arcsec, 2.2\arcsec\ ). The contours are at 1$\sigma$ above the noise level. The pink dashed box denotes the region used to calculate the magnetic flux and magnetic energy for the QSEB, as shown in Fig.~\ref{fig:10}.}
    \label{fig:9}
\end{figure*}

\section{Results}
\label{sec:results}
In this study, we highlight four different magnetic topologies that can be associated with QSEBs and their related UV brightenings. 
Figure \ref{fig:2} presents simple cartoon sketches of these four topologies, which can be briefly described as follows:

\begin{enumerate}[label=\alph*)]
  \item The QSEB and UV brightening occur at the polarity inversion line between a dipolar arrangement of opposite polarities.
\end{enumerate}
The UV brightening occurs at a 3D magnetic null with a fan-spine configuration and
\begin{enumerate}[label=\alph*), resume]
  \item the QSEB occurs around the footpoint of the outer spine.
  \item the QSEB occurs around the footpoint of the inner spine.
  \item a QSEB or multiple QSEBs occur around the footpoints of the fan surface.
\end{enumerate}
We present observations corresponding to each of these configurations in the subsequent subsections.

\subsection{QSEBs and UV brightening with a dipole configuration}  \label{sec: 4.1}

We observe two QSEBs in Region~1, which have a dipole magnetic topology. Panel a) of Fig.~\ref{fig:3} shows the flame-like QSEB in the blue wing of the \Hbeta\ line.
We discuss two instances of QSEB activity - i) QSEB-1 from 08:56:22 UT with a duration of 202~s, and ii) QSEB-2 from 09:03:57 UT with a duration of 87~s. 
The snapshot in Fig.~\ref{fig:3} at time 09:04:47 UT corresponds to the QSEB-2 event when the QSEB exhibits maximum wing enhancement. 
Significant increases in wing intensity are observed at other times as well, which can be seen in the accompanying movie. 
The snapshot shows that the QSEB-2 occurred on the stronger positive polarity, as shown in panel b). 
However, the corresponding movie (see online version) reveals that during the first event, QSEB-1 starts between the two opposite polarities and slowly moves to the stronger positive patch during the second event. 
The location for which the \Hbeta\ spectra are displayed, indicated by the blue marker over the QSEB, was adjusted a few times to sample the strongest QSEB spectra throughout the sequence. 
The SJI~1400 image in Panel d) of Fig.~\ref{fig:3} shows a UV brightening near the QSEB with $>$5$\sigma$ contours. These events appear before, after, and during the occurrence of the two QSEBs.  
Fig.~\ref{fig:4} illustrates the magnetic topology near the QSEBs at various timesteps, which is similar to the sketch in panel a) of Fig.~\ref{fig:2}, where the field lines connect the two opposite polarities. 
Panel a) shows the topology before the onset of QSEB-1 at 08:56:15 UT, with extreme $B_\mathrm{LOS}$ values of $-$49 and $+$70~G at the loop footpoints. 
Panel b) depicts the topology when the QSEB occurs at 08:56:58 UT, with extreme $B_\mathrm{LOS}$ values of $-$36 and $+$68~G at the footpoints of the dipole.
Similarly, for QSEB-2, panel c) denotes the instance before its occurrence at 09:03:07 UT, with extreme $B_\mathrm{LOS}$ values of $-$100 and $+$93~G at the dipole footpoints. 
These values reduce to extreme $B_\mathrm{LOS}$ of $-$64 and $+$77~G when the QSEB-2 occurs at 09:05:02 UT, shown in panel d).
We observe slight flux emergence before the onset of QSEB-1 from 08:53:29 UT.
The changes in magnetic flux and magnetic energy are shown in Fig.~\ref{fig:5} a) and b). 
For calculating the magnetic flux, we consider the region enclosed by the pink dashed box in Fig.~\ref{fig:3} b). 
For the magnetic energy of the potential field, we consider a small volume around the QSEB, with the same bottom boundary as used for calculating the magnetic flux and extending slightly beyond the height of the loops in the $z$-direction (till $z$ = 415~km).
The apex height of the loops at 08:56:15 UT, just before QSEB-1 begins, is 200~km above the photosphere. 
The cancellation between the two opposite polarities begins at 08:56:30 UT, reducing the magnetic energy within the region.
The magnetic energy before the onset of this QSEB is $7.27 \times 10^{24}$~ergs, which decreases by $7.06 \times 10^{23}$~ergs by the time the QSEB ends.
The loop apex height also starts to decrease, reaching 90~km above the photosphere at 08:56:58 UT, when QSEB-1 occurs. 
After QSEB-1 ends, a considerable flux emergence is observed in the region, resulting in a build-up of magnetic energy in the volume to $1.71 \times 10^{25}$~ergs by 09:03:50 UT. 
Consequently, the loop height increases to 323~km before the beginning of QSEB-2. 
This QSEB initiates at 09:03:57 UT as soon as the magnetic energy begins to drop. 
The loop height also reduces to 160~km during the strongest enhancements of the \Hbeta\ line. 
The magnetic energy continues to decrease until 09:06:51 UT, with a total of $9.26 \times 10^{24}$~ergs being released. 
The corresponding $B_\mathrm{LOS}$ values for both the cases mentioned above show that the loop height is proportional to the $B_\mathrm{LOS}$ at the footpoints of the loops. 
The heights of these loops increase when there is flux emergence and decrease when there is cancellation.
We see that more magnetic energy is released when QSEB-2 occurs, indicating more cancellation, which has stronger $B_\mathrm{LOS}$ at the loop footpoints compared to QSEB-1. 
The accompanying movie also shows that QSEB-2 shows more wing enhancement of the \Hbeta\ spectra than QSEB-1. 
The UV brightening is shown close to the loop tops of the QSEB with some spatial offset towards the direction of the limb, which is shown in Fig.~\ref{fig:6}.

\subsection{UV brightening at 3D null, QSEB at outer spine footpoint}  
\label{sec: 4.2}

Figure \ref{fig:7} a) and b) shows an observation corresponding to the topology illustrated in Fig.~\ref{fig:2}b. 
The UV brightening, shown as a yellow region, occurs near the 3D null point, which is located 193~km above the photosphere. 
Two QSEBs are observed in close proximity to this UV brightening; we call them QSEB-A and QSEB-B. 
QSEB-A is located where the outer spine of the 3D null connects with the positive polarity. 
In the snapshot shown in Fig.~\ref{fig:7} at 08:21:20 UT, it has a magnetic bright point type spectrum with no EB-like wing enhancements. 
However, after 90~s, this magnetic bright point evolves into a QSEB, displaying strong intensity enhancements in the wings of the \Hbeta\ line. 
The QSEB-A occurs on the positive patch, which has extreme $B_\mathrm{LOS}$ value of $+$89~G before the QSEB begins and increases to $B_\mathrm{LOS}$ of $+$148~G during the QSEB event.
In Fig.~\ref{fig:8} a) and b), we see flux emergence and increase in the magnetic energy begin right when this QSEB occurs, indicated by the yellow-shaded region.  
The reconnection at the 3D null point causes localised heating, which we observe as the UV brightening near the null point. 
QSEB-A likely occurs due to energy transport from the reconnection site at the 3D null along the outer spine where the field is strong. 
The magnetic field is very weak at the footpoint of the inner spine, which could explain why there is not sufficient heating to produce an observable QSEB.
QSEB-B is also observed close to the UV brightening, but based on the magnetic connectivity, it does not appear to be triggered by the same magnetic reconnection responsible for the UV brightening. 
This QSEB exhibits a dipole configuration, similar to Fig.~\ref{fig:2} a). 
The footpoints of the loops connecting the two polarities, along with the footpoints of the inner spine, the fan surface, and the outer spine of the 3D null, are depicted in Fig.~\ref{fig:7} c). 
The evolution of magnetic flux shows the cancellation of opposite polarities during its occurrence, as shown in Fig.~\ref{fig:8} c). 
The magnetic energy for both QSEBs is calculated within a box around the QSEBs with the bottom boundary as the pink dashed boxes shown in Fig.~\ref{fig:7} c), and considering height till $z$ = 276~km. 
The magnetic energy for QSEB-B increases to $2.06 \times 10^{24}$~ergs and drops by $1.00 \times 10^{24}$~ergs by the time the QSEB ends.
While the QSEB-B and the UV brightening share a common negative polarity, their respective reconnections occur on opposite sides of this polarity. 
As a result, this QSEB and UV brightening are not directly connected to each other.

\begin{figure}
    \centering
    \includegraphics[width=\linewidth]{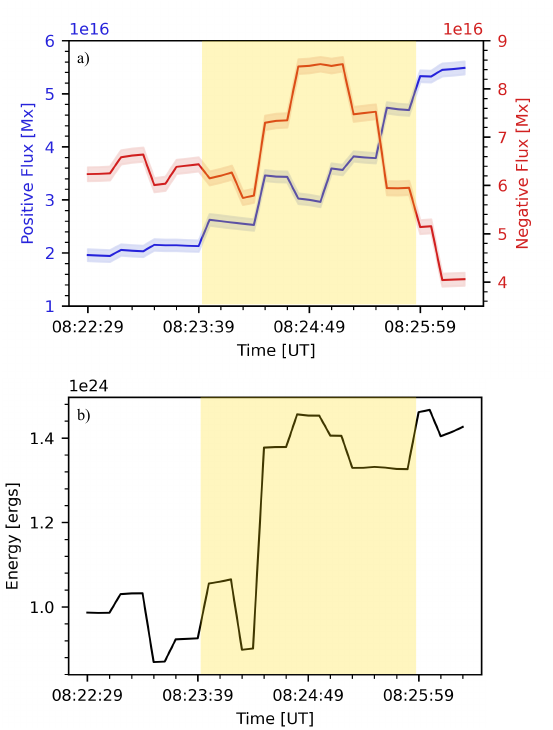}
    \caption{Same as in Fig.~\ref{fig:5}, showing the evolution of magnetic flux and magnetic energy for the QSEB in Fig.~\ref{fig:9}. The error in positive and negative flux is shown as thin-shaded regions along the curves.}
    \label{fig:10}
\end{figure}

\subsection{UV brightening at 3D null, QSEB at footpoints of the fan surface} 
\label{sec: 4.3}
In section \ref{sec: 4.2}, we presented a QSEB close to the UV brightening with a dipole configuration at 08:21:20 UT. 
Four minutes later, at 08:25:10 UT, a QSEB is observed again at the same location. 
This time, the fan surface/dome associated with the 3D null has expanded, with some of its footpoints now situated at the positive polarity where the QSEB occurs, as shown in Fig.~\ref{fig:9} a).
A sketch of this topology is shown in panel d) of Fig.~\ref{fig:2}.
In the time series that follows, we observe many QSEBs near the footpoints of this dome structure. 
The inner spine is rooted in a negative polarity patch, while the footpoints of the dome connect to nearby positive polarities. 
These QSEBs typically occur at the stronger positive polarity regions linked to the footpoints of the fan surface of the 3D null. 
Here, we present a case where the QSEB occurs at one of these stronger positive polarities, which has extreme $B_\mathrm{LOS}$ value of $+$54~G, though, in the subsequent time series, QSEBs are also observed at two other strong positive polarity patches.
The QSEB could be happening due to energy flow from the reconnection site to the stronger polarities, causing localised heating.
The 3D null is positioned at a height of approximately 480~km above the photosphere, rooted in a negative polarity patch, which has an extreme $B_\mathrm{LOS}$ value of $-$45~G during this event. 
The UV brightening occurs close to the 3D null point, where reconnection likely occurs due to misalignment between magnetic field lines of the fan surface and the emerging dipole loops, potentially forming quasi-separatrix layers and current sheets.
Panel a) of Fig.~\ref{fig:9} shows the dipole configuration of this QSEB, while panel b) shows the $B_\mathrm{LOS}$ map with one of the footpoints of the dipole being located on the same negative polarity patch where the inner spine of the 3D null is located. 
This indicates that a common reconnection could be the source of both the QSEB and the UV brightening. 
Figure \ref{fig:10} a) shows flux emergence, with the positive polarity at the QSEB location increasing its flux even while the QSEB occurs. 
Only the negative polarity, where the footpoints of the dipole of the QSEB are located, shows some decrease in flux from 08:25:17 UT. 
For this calculation, we have considered a box around the QSEB, with the bottom boundary as the pink dashed region shown in Fig.~\ref{fig:9} and taking height till $z$ = 415~km.
This type of configuration seems to be a miniature version of the larger topologies studied in numerical simulations, where the emerging dipoles are shown to affect the preexisting null point configuration like in coronal bright points \citep{2022ApJ...935L..21N} or flux rope eruptions \citep{2024ApJ...962..149T}.

\begin{figure*}
    \centering
    \includegraphics[width=\linewidth]{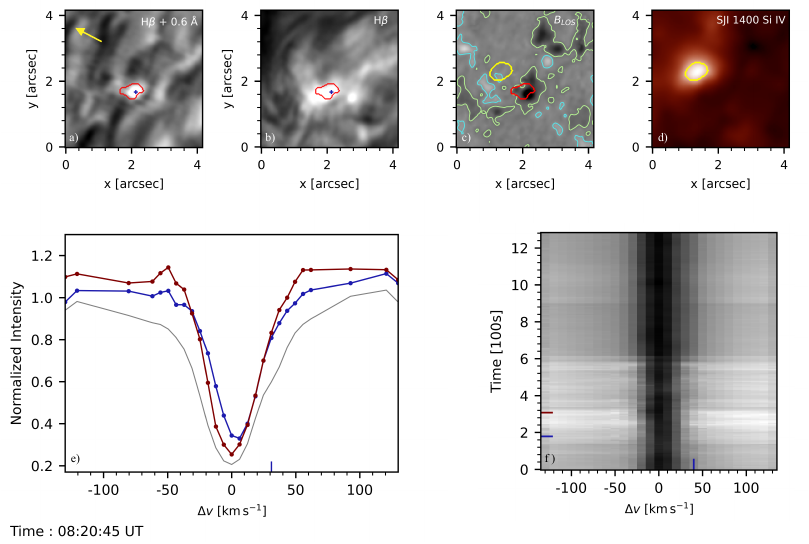}
    \caption{Details of the QSEB occurring at the footpoint of the inner spine of a 3D fan-spine topology in Region~2. The top row shows the QSEB in \Hbeta\ $+0.6$~\AA, in \Hbeta\ core, the $B_\mathrm{LOS}$ map with contours at 2$\sigma$ above the noise level, and SJI 1400. The yellow contours in panel d) show region with $>$5$\sigma$ intensity in SJI~1400. Panel e) shows the normalised intensity in \Hbeta\ at the blue cross in the above \Hbeta\ $+0.6$~\AA\ image. The \Hbeta\ spectra are shown for two instances - blue shows significant line core brightening, and maroon shows wing enhancement. The \Hbeta\ core also shows the QSEB, as this event has considerable brightening of the line core.  Panel f) shows the $\lambda t$ diagram for the blue marker over the QSEB. The position of this marker has been adjusted a few times to obtain the strongest spectra for each timestep. The vertical blue markers in panels e) and f) denote the position of the wing of \Hbeta\ line for which the image is shown in panel a). The horizontal blue marker in panel f) denotes the time of the snapshot shown in this figure. The horizontal maroon marker denotes the time for the maroon \Hbeta\ spectra in panel e). The yellow arrow shows the direction to the nearest limb. An animation of this figure, which shows the evolution of this QSEB, is available in the online material (see \url{http://tsih3.uio.no/lapalma/subl/qseb_uvb_topology/movies/bhatnagar_movie_fig11.mp4}).}
    \label{fig:11}
\end{figure*}

\begin{figure*}
    \centering
    \includegraphics[width=\linewidth]{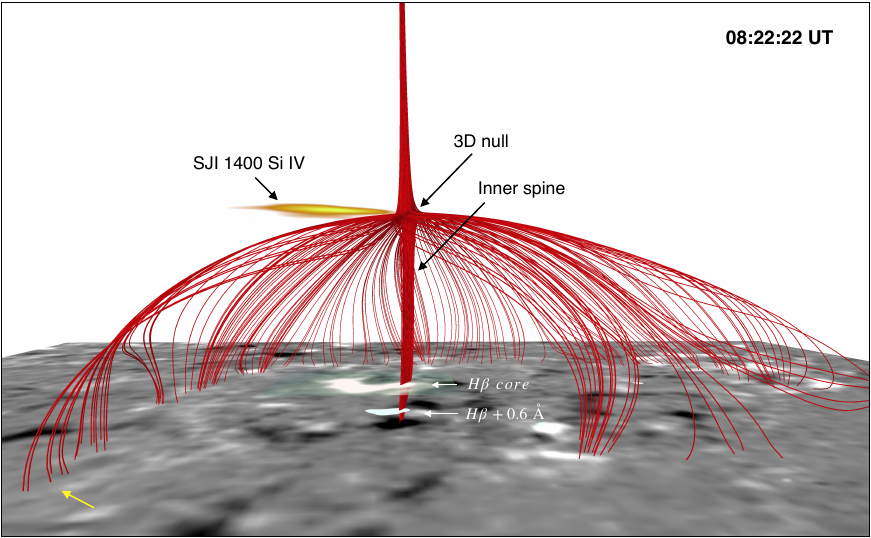}
    \caption{Magnetic fan-spine topology showing the UV brightening at the 3D null and QSEB at the footpoint of the inner spine. This resembles the sketch shown in Fig.~\ref{fig:2} c). The inner spine is situated on negative polarity, and the dome footpoints connect to the nearby positive polarities. The height of the 3D null is 2.3~Mm above the photosphere. The QSEB is shown in both \Hbeta\ $+0.6$~\AA\ and in the \Hbeta\ core. The yellow arrow shows the direction to the nearest limb.}
    \label{fig:12}
\end{figure*}

\begin{figure*}
    \centering
    \includegraphics[width=\linewidth]{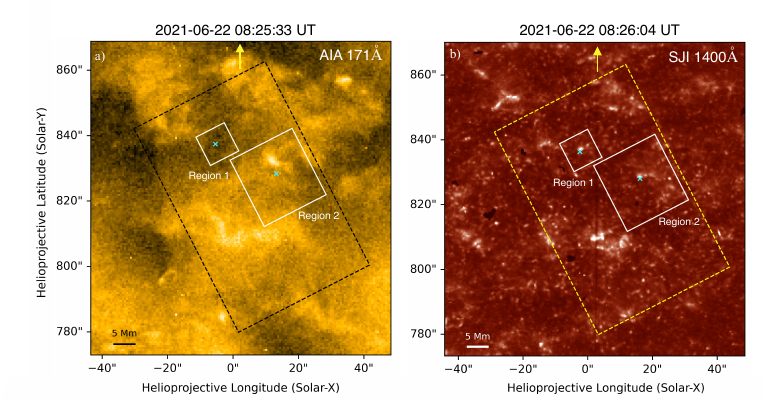}
    \caption{Overview of the observed SST region in AIA 171 \AA\ and SJI 1400 \AA, at their original resolution. The black dashed box in AIA 171 \AA\ and yellow dashed box in SJI 1400 \AA\ outline the CHROMIS field of view. White boxes highlight Region~1 and Region~2, with cyan cross markers indicating the locations of the QSEBs. Both QSEBs are associated with nearby brightenings in SJI 1400 \AA, but only Region~2 shows a corresponding brightening in AIA 171 \AA\ close to the QSEB. The yellow arrow points toward the north solar limb.}
    \label{fig:13}
\end{figure*}
\subsection{UV brightening at 3D null, QSEB at inner spine footpoint}
\label{sec: 4.4}
In Region~2, we identify a QSEB whose topology corresponds to Panel c) of the cartoon in Fig.~\ref{fig:2}. 
The QSEB is shown in the red wing and the core of \Hbeta\ line in Fig.~\ref{fig:11}.
There are two events of this QSEB - i) from 08:18:54 UT to 08:27:18 UT, with a duration of 504~s and ii) from 08:33:00 UT to 08:34:26 UT, with a duration of 86~s.
Throughout most of its duration, the QSEB exhibits brightening primarily in the \Hbeta\ line core spectra, with intermittent wing enhancements.
We show \Hbeta\ line spectra of two instances in panel e), one with significant line core brightening and another with wing enhancement.
Even after the QSEB has ended, brightening in the \Hbeta\ line core persists. 
A corresponding movie for Fig.~\ref{fig:12} is available as online material that further demonstrates this.
The snapshot in Fig.~\ref{fig:11} and the accompanying movie also show a UV brightening near the QSEB, with a spatial offset toward the limb direction. 
The UV brightening overlaps temporally with the QSEB for most of its duration, occurring from 08:19:15 UT to 08:23:46 UT and again from 08:25:45 UT to 08:26:56 UT, which coincides with the first QSEB event. 
Another UV brightening event occurs from 08:28:15 UT to 08:30:02 UT, just after the first QSEB detection has ended, although enhanced \Hbeta\ line core brightening persists during this period. 
During the second QSEB event, the UV brightening is detected above the 5$\sigma$ threshold for a couple of instances, but the \Siiv\ region appears bright near the QSEB.

Figure \ref{fig:12} illustrates the magnetic topology linking the QSEB and the UV brightening. 
The UV brightening occurs close to the 3D null of this topology, while the QSEB occurs at the footpoint of the inner spine, suggesting a connection between the two phenomena.
The inner spine of the 3D null remains anchored at the negative polarity patch for a total duration of about 21~min, even after the QSEB has ended. 
The strength of the $B_\mathrm{LOS}$ varies between $-$93~G and $-$220~G at the footpoint of the inner spine during this duration.
The footpoints of the dome continue to connect to the nearby positive polarities.
During this period, the height of the 3D null varies between 2.2~Mm to 2.6~Mm due to variations in $B_\mathrm{LOS}$.
Both the QSEB events primarily occur on the negative polarity patch throughout their lifetime, as shown in panel c) of Fig.~\ref{fig:11}. 
A few weak opposite polarity patches are also present near this negative patch, where reconnection could cause the QSEB. 
However, given the stability of the 3D null point structure, with the inner spine footpoint anchored at the QSEB location for the entire duration, it is possible that the QSEB results from the energy transfer along the inner spine after the reconnection at the 3D null point. 
This would explain the UV brightening happening before the wing enhancements, with initial heating observed in the \Hbeta\ line core, followed by intermittent wing enhancements as the event progresses.
In this region, we also observe a brightening in the AIA 171 \AA\ channel, located 2--4~Mm towards the limb from the QSEB throughout the sequence. 
Since there are no significant magnetic polarities near the AIA 171 \AA\ brightening, aside from the polarities where the QSEB is observed, this brightening may be related to reconnection happening at the null point, whose inner spine footpoint is at the QSEB. 
A map of the AIA 171 \AA\ channel in Fig.~\ref{fig:13} shows the brightening in Region~2, close to the cyan marker representing the QSEB.
Similar brightening is also observed in other AIA channels, such as 304 \AA\ and 193 \AA\ (not shown in the figure).
%

\section{Discussion}
\label{sec:discussion}
This study investigates the magnetic topologies associated with co-spatial and co-temporal QSEBs and UV brightenings.
QSEBs were identified from the \Hbeta\ data using a \textit{k}-means clustering algorithm. 
High-resolution photospheric magnetograms from the SST enabled us to resolve the magnetic structures underlying these events, while IRIS observations provided information about their transition region response.
Through potential field extrapolations applied to the SST $B_\mathrm{LOS}$ data, we identified four distinct magnetic topologies that connect QSEBs and UV brightenings, ranging from simple dipole configurations to more complex fan-spine structures, each associated with different reconnection scenarios. 
By examining the magnetic topology of QSEBs and UV brightenings together, we provide insights into the reconnection processes driving these events.
In the following subsections, we mention the limitations of the current study, discuss the topologies and the summary and conclusion.

\subsection{Limitations}
Here, we discuss some of the limitations of the current methodology and their impact on the results of our analysis. 
This study focuses on a quiet Sun region where magnetic fields are quite weak. 
This region is close to the North limb, which adds substantial projection effects when comparing events at different heights.
This resulted in noisy measurements of the local transverse magnetic field components ($B_\mathrm{x}$ and $B_\mathrm{y}$), which are inferred from the Stokes Q and U profiles. 
Due to this, it was difficult to resolve the 180$^{\circ}$ ambiguity and perform reliable vector transformations, as it added substantial noise to the vector magnetic field.
Hence, we only used the line-of-sight magnetic field ($B_\mathrm{LOS}$, derived from Stokes V) for the magnetic field extrapolations and performed the potential field extrapolation for this work without correcting for any projection effects. A potential field extrapolation assumes no electric currents are present in the region. 
However, this approximation might be reasonable for quiet Sun regions studied here, where we expect relatively small currents due to weaker magnetic field strengths.
As the potential field is the minimum energy configuration for a given bottom boundary, using potential field extrapolations does not give us information on the free magnetic energy. 
Hence, we could only approximately estimate the energy released during the event based on the changes in the potential field energy.
Although we did not use any specific null-detection algorithm, our analysis consists of finding the null points, which are robust topological features. 
It has been shown by \citet{2009SoPh..254...51L} that potential field extrapolations can be used to find these accurately.

\subsection{Discussion of the topologies}
In section \ref{sec: 4.1}, we presented the dipole configuration, which is often associated with various scenarios for reconnection, like U-shaped photospheric magnetic loops and bald patches \citep{2004ApJ...614.1099P, 2006AdSpR..38..902P, 2012ASPC..455..177P, 2012EAS....55..115P}. 
In this case, QSEBs occur between regions of opposite magnetic polarity near the PIL, a well-established scenario for EBs, as seen in previous studies. 
QSEBs in dipole configurations in our study show magnetic flux cancellation between opposite polarities, consistent with earlier findings for EBs in active regions by, for example, \citet{2013ApJ...774...32V} and \citet{2016ApJ...823..110R}. 
We observe QSEBs either at the PIL between two opposite polarities or slightly offset from the PIL and located on the stronger magnetic polarity patch. 
The presence of QSEBs near stronger polarity patches, alongside adjacent opposite polarities, aligns with the results of \citet{2016A&A...592A.100R}.
\citet{2002ApJ...575..506G} report energy release for EBs in the range of $10^{26}$ to $10^{28}$~ergs, while \citet{2016ApJ...823..110R} found slightly lower values, between $10^{23}$ and $10^{26}$~ergs. 
In comparison, the energies we observe for QSEBs fall within the lower limit of these estimates, ranging from $10^{23}$ to $10^{24}$~ergs, as QSEBs represent smaller-scale reconnection events and have weaker magnetic fields as compared to EBs.

In sections \ref{sec: 4.2} -- \ref{sec: 4.4}, we observe UV brightenings predominantly near the 3D null points with a fan-spine configuration, in line with the findings of \citet{2017A&A...605A..49C} and \citet{2018A&A...617A.128S} who show fan-spine topology as key sites for UV bursts. 
The UV brightenings studied here do not have the extreme spectral profiles of UV bursts. 
Some examples of spectra of our UV brightenings are shown in \citetalias{2024A&A...689A.156B}, which have lower intensities than UV bursts. 
These do not show the broad emissions and absorption features in the \Siiv\ 1394 \AA\ line that are characteristic of UV bursts. 
Therefore, they may not always be direct signatures of magnetic reconnection but consequences of energy transfer associated with the reconnection event. 
It is also possible that the events are not energetic enough to produce strong emission.
\citet{2017A&A...605A..49C} reported the height of the 3D null point to be 0.5~Mm above the photosphere, where magnetic reconnection took place. 
In our observations, the height of the 3D null points varies considerably, ranging from 0.2~Mm to 2.6~Mm, depending on the region and the magnitude of the $B_\mathrm{LOS}$ values at the footpoint of the inner spine, suggesting a broader range of reconnection altitudes within the fan-spine configuration. 
Region~1 has null points between 0.2~Mm and 0.5~Mm, with no brightenings visible in the AIA 171 \AA\ channel. 
This may result from the lower height of the 3D null due to weaker magnetic field strength at the footpoints, limiting energy release and reducing the likelihood of showing any coronal signatures. 
In contrast, for Region~2, the magnetic field strength at the footpoints of the null point is stronger compared to Region~1. 
As a result, the null points are formed in the upper transition region/lower corona (2.2 to 2.6~Mm), and continuous brightening is observed in the AIA 171 \AA\ channel, indicating that the reconnection may occur at higher altitudes, driving activity in the lower corona. 
The observed AIA 171 \AA\ brightening appears diffuse and is not directly located above any strong magnetic fields. 
The magnetic patch at which the QSEB is located is the closest magnetic field concentration near the AIA 171 \AA\ brightening. 
This could be similar to the campfires described by \citet{2021A&A...656L...4B}, which appear fuzzy in AIA 171 \AA, 304 \AA, 193 \AA\ channels with spatial scales of 0.4~Mm to 4~Mm and heights between 1~Mm and 5~Mm. 
\citet{2022A&A...660A.143K} also found that 11 out of 38 of their campfires were located far from magnetic footpoints, with the nearest footpoint approximately 2~Mm away, in quiet Sun observations near the disc centre. 
In our case, the reconnection at the null point likely causes heating across multiple heights channelled by the inner and outer spine. 
We observe a QSEB at the footpoint of the inner spine, UV brightenings in SJI 1400, which are slightly offset from the null point, and AIA brightenings in the lower corona, which show a greater offset from the QSEB. 
Since our observation is close to the north solar limb, there is a significant projection effect ($\mu=0.48$ corresponding to a viewing angle of 61$^{\circ}$). 
For every 1~Mm of height in the solar atmosphere, this geometry introduces an offset of approximately 1.78~Mm in the observed position. 
Such projection effects can cause offsets in the aligned datasets between diagnostics that are formed at different heights.  
Given that the null points in Region~2 are between 2.2~Mm and 2.6~Mm, this projection would cause a displacement of around 3.92~Mm to 4.64~Mm, aligning well with the observed AIA 171~\AA\ brightening being 2 to 4~Mm away from the QSEB.

We identify three magnetic configurations for QSEBs involving fan-spine structures, each linked to UV brightenings at the 3D null point and QSEBs at the outer spine, inner spine, and dome footpoints. 
While QSEBs may result from photospheric reconnection, their proximity to UV brightenings suggests a connection to the larger fan-spine structure, with energy transfer likely driven by reconnection at the 3D null point or at quasi-separatrix layers that may form between the fan surface and the emerging dipole where the QSEB occurs.
These scenarios resemble large-scale reconnection events seen in solar flares, where simultaneous brightenings are observed across these different magnetic footpoints. 
In our observations, the QSEB at the inner spine footpoint begins alongside the appearance of the 3D null point. 
Similarly, in solar flares, brightenings at inner spine footpoints occur almost simultaneously with reconnection at the 3D null, as seen in \citet{2017ApJ...837..173R}. 
Furthermore, studies of circular ribbon flares such as \citet{2020ApJ...903..129P, 2024A&A...687A.172J, 2024RvMPP...8....7Z} highlight that in these flares, basically, all footpoints of the fan surface brighten as continuous circular ribbon structures. In contrast, QSEBs usually appear at those footpoints of the dome where the magnetic field is strongest.
In the case of QSEBs at the outer spine footpoint, energy may propagate from the reconnection site, similar to the scenario of flares described by \citet{2012A&A...547A..52R} and \citet{2013ApJ...769..112D}. 
It is possible that the outer spine could be open or connected to a region that lies outside the observed region.
In addition to these scenarios, QSEBs could also result from emerging flux at the polarity where they occur, which cancels with the overlying magnetic topology. 
For solar flares, these brightenings at all footpoints are mostly observed together. 
However, we do not observe all these scenarios occurring together for QSEBs. 
Unlike solar flares, QSEBs are much smaller-scale reconnection events occurring lower in the atmosphere. 
It could be that in these smaller events, QSEBs occur only at those footpoints where the magnetic field strength is strongest. 

\subsection{Summary and Conclusion}
To summarise, in this paper, we have presented examples of observations related to various magnetic topologies associated with QSEBs and UV brightenings. 
We find that QSEBs can occur due to magnetic reconnection between opposite polarities in a simple dipole configuration. However, they can also occur as part of complex magnetic topologies involving 3D nulls with fan-spine structures. The reconnection can happen higher in the atmosphere, with the energy being transferred down to cause the temperature increase needed to observe the wing enhancement in the \Hbeta\ spectral line. It is important to note that multiple scenarios presented likely occur simultaneously, which results in a combination of different magnetic topologies. The scenarios discussed in this paper also explain the co-temporal and co-spatial occurrences of QSEBs and UV brightenings. 

In a follow-up paper, we will explore these fan-spine topologies further, with QSEBs occurring at multiple footpoints of the dome and accompanied by inverted Y-shaped jets and UV brightenings.
For future studies, it would be beneficial to use observations which are closer to the disc centre, as this would reduce the projection effects. Having $\mu$ angles slightly greater than the observations used in this work will give better magnetic field measurements and also allow for observing the QSEB flames. Additionally, similar topological studies can be done for EBs and UV bursts in active regions, which involve reconnection between stronger magnetic fields.
\begin{acknowledgements}
We thank Guillaume Aulanier for the helpful discussions and Reetika Joshi for aligning the AIA datasets to SST data. 
The Swedish 1-m Solar Telescope (SST) is operated on the island of La Palma by the Institute for Solar Physics of Stockholm University in the Spanish Observatorio del Roque de los Muchachos of the Instituto de Astrof{\'\i}sica de Canarias.
The SST is co-funded by the Swedish Research Council as a national research infrastructure (registration number 4.3-2021-00169).
This research is supported by the Research Council of Norway, project number 325491, 
and through its Centres of Excellence scheme, project number 262622. 
IRIS is a NASA small explorer mission developed and operated by LMSAL, with mission operations executed at NASA Ames Research Center and major contributions to downlink communications funded by ESA and the Norwegian Space Agency.
SDO observations are courtesy of NASA/SDO and the AIA science teams.
%
J.J. is grateful for travel support under the International Rosseland Visitor Programme. 
J.J. acknowledges funding support from the SERB-CRG grant (CRG/2023/007464) provided by the Anusandhan National Research Foundation, India.
A.P. and D.N.S acknowledge support from the European Research Council through the
Synergy Grant number 810218 (``The Whole Sun'', ERC-2018-SyG).
We acknowledge using the visualisation software VAPOR (www.vapor.ucar.edu) for generating relevant graphics.
We made much use of NASA's Astrophysics Data System Bibliographic Services.
\end{acknowledgements}

\bibliographystyle{aa}
\bibliography{aditis_ref} 

\end{document}